\documentclass[prb,showpacs,twocolumn,superscriptaddress,floatfix]{revtex4}
\usepackage[english]{babel}
\usepackage{graphicx}
\usepackage{amsmath}
\usepackage{amssymb}
\usepackage{epsfig}
\begin{document}
\title{Splitting of Andreev levels in a Josephson junction by spin-orbit coupling}
\author{B. B{\'e}ri}
\affiliation{Department of Physics of Complex Systems,
E{\"o}tv{\"o}s University,
 H-1117 Budapest, P\'azm\'any P{\'e}ter s{\'e}t\'any 1/A, Hungary}
\author{J.~H.~Bardarson}
\affiliation{Instituut-Lorentz, Universiteit Leiden, P.O. Box 9506,
2300 RA Leiden, The Netherlands}
\author{C.~W.~J.~Beenakker}
\affiliation{Instituut-Lorentz, Universiteit Leiden, P.O. Box 9506,
2300 RA Leiden, The Netherlands}
\date{September, 2007}
\begin{abstract}
We consider the effect of spin-orbit coupling on the energy levels of a
 single-channel  Josephson junction below the superconducting gap. We investigate
quantitatively the level splitting arising from the combined effect of  spin-orbit
coupling and the time-reversal symmetry breaking by the phase difference between the superconductors. Using the scattering matrix approach we establish a simple
connection between the quantum mechanical time delay matrix 
and the effective Hamiltonian for the level splitting. As an
application we calculate the distribution of level splittings for an
ensemble of chaotic Josephson junctions.
The distribution falls off as a power law for large splittings, unlike the
exponentially decaying splitting distribution given by the Wigner surmise --
 which applies for normal chaotic quantum dots with spin-orbit coupling in 
the case that the time-reversal symmetry breaking is due to  a magnetic field.
\end{abstract}
\pacs{74.45.+c, 71.70.Ej, 05.45.Pq, 74.78.Na}
\maketitle{}

\section{Introduction}
\begin{figure}[tb]
  \begin{center}
    \includegraphics[width=0.8\columnwidth]{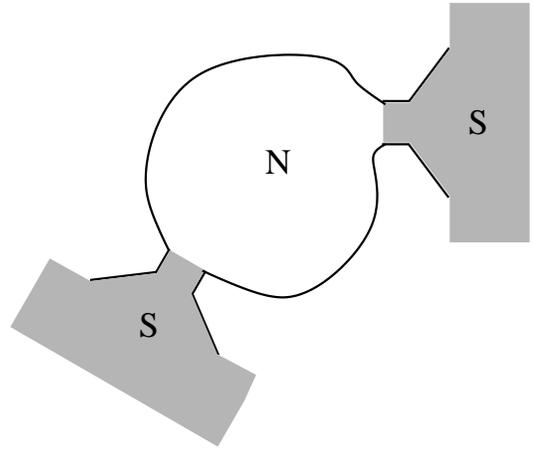} 
  \end{center}
  \caption{Sketch of a quantum dot Josephson junction: the quantum dot (N)  is
  connected to  two 
  superconductors (S) by  point contacts. Spin-orbit coupling splits the energy levels
  of the system
 when the superconductors have a nonzero phase difference.}
  \label{fig:setup}
\end{figure}

A Josephson junction is a weak link between two superconductors with an
adjustable phase difference $\phi$. The weak link may be a tunnel barrier or a
normal metal. Fig.~\ref{fig:setup} shows, for example, a Josephson junction
consisting of a small piece of normal metal (a quantum dot), connected to the
superconductors by a pair of narrow constrictions (quantum point contacts). 
The excitation spectrum below the superconducting gap $\Delta$ consists of
discrete energies, called Andreev levels. 
In zero magnetic field, the energy levels $\varepsilon_n$ are determined by
the normal-state transmission eigenvalues $T_n$ if $\Delta \ll
\hbar/\tau_{\rm dw}$, where $\tau_{\rm dw}$ is the dwell time of an electron
in the normal region (before it is converted into a hole by Andreev reflection
at the superconductor). The relationship is\cite{Bee91}
\begin{equation}
\varepsilon_{n}=\Delta \sqrt{1-T_n  \sin^2(\phi/2)}+{\cal O}(\Delta^2\tau_{\rm
  dw}/\hbar).\label{eq:TnEnrel}
\end{equation}
Each level is twofold spin-degenerate (Andreev doublet).

Recently the effect of spin-orbit coupling on
Josephson junctions became a subject of investigation\cite{Bez02,Cht03,Krive04,Dim06,DZEM07}. 
This is a subtle effect for the following reason: 
On the one hand, in the absence of magnetic fields the normal-state
transmission eigenvalues $T_n$
are Kramers degenerate because of the time-reversal invariance of the normal
system. On the other hand, one would expect a breaking of the degeneracy of
the Andreev doublets
 because the phase difference between the superconducting
contacts breaks the time-reversal symmetry of the system. Still, to leading
order in $\Delta\tau_{\rm dw}/\hbar$  the one-to-one relationship
\eqref{eq:TnEnrel} between $\varepsilon_n$ and $T_n$ ensures that the Andreev levels remain degenerate for nonzero $\phi$. As was pointed out by Chtchelkatchev and
Nazarov\cite{Cht03}, to see a splitting of the Andreev doublets
 as a result of the combined effect of spin-rotation symmetry
breaking by spin-orbit coupling and time-reversal symmetry breaking by the
phase difference one has to go beyond the leading order in $\Delta\tau_{\rm
  dw}/\hbar$.
This tunable level splitting was exploited in a proposal of
Andreev qubits for quantum computation\cite{Cht03}.

In this work we examine the splitting of the Andreev doublets
quantitatively by calculating the first order correction to the energy levels
in the small parameter $\Delta\tau_{\rm dw}/\hbar$. We concentrate our
attention on the case when the quantum point contacts support one propagating mode each. We give a simple relation between the
effective Hamiltonian for the level splitting of Chtchelkatchev and
Nazarov\cite{Cht03} and the Wigner-Smith time delay matrix, 
\begin{equation}Q=-i S^\dagger \frac{dS}{d\varepsilon},\end{equation}
 where $S$ is the scattering matrix of the normal system. 
As an application, we calculate how the splittings are distributed for an ensemble
of systems where the two superconductors are connected by a chaotic quantum
dot, assuming that the spin-orbit coupling in the dot is strong enough that the
dot Hamiltonian can be modeled as a member of the symplectic ensemble of Random Matrix Theory
(RMT)\cite{mehta,Bee97}. 
The present study in the regime $\Delta \ll \hbar/\tau_{\rm dw}$ complements
earlier work\cite{Alt97,BBB07} in the opposite regime $\Delta \gg \hbar/\tau_{\rm dw}$.

The paper is organized as follows. In Sec.~\ref{sec:SQsplit} we employ the
scattering matrix approach for calculating the first order correction in
$\Delta \tau_{\rm dw}/\hbar$ to the
Andreev levels, and obtain the effective Hamiltonian for the level splitting
in terms of  the time delay matrix $Q$. 
For simplicity, we consider the single-channel case in Sec.~\ref{sec:SQsplit}
and give the multichannel extension in an Appendix. We apply our
single-channel formula to a calculation of the splitting distribution for an
ensemble of chaotic Josephson junctions in Sec.~\ref{sec:chQdot}. 
We conclude in Sec.~\ref{sec:conclude} with a
comparison of the splitting distribution of the Andreev doublets and the Wigner surmise of RMT.

\section{Splitting Hamiltonian and Wigner-Smith matrix}
\label{sec:SQsplit}
For energies below the superconducting gap $\Delta$ the Josephson junction supports bound
states, with excitation energies given by the roots of the secular equation\cite{Bee91}
\begin{eqnarray}
{\rm Det}\,\left[\openone-\alpha(\varepsilon)^2
r_{\rm A}^{\ast}S_{\rm e}(\varepsilon)r_{\rm A}
S_{\rm h}(\varepsilon)\right] =0,\label{eq:det4spect}
\end{eqnarray}
where 
\begin{eqnarray}
\alpha=\exp\left[-{\rm i}\arccos\left(\frac{\varepsilon}{\Delta}\right)\right],\,
r_{\rm A}=
{\renewcommand{\arraystretch}{0.6}
\left(\begin{array}{cc}
{\rm e}^{{\rm i}\phi/2}\openone&0\\
0&{\rm e}^{-{\rm i}\phi/2}\openone
\end{array}\right)},
\label{eq:alpharA}
\end{eqnarray}
and $S_{\rm e}(\varepsilon)$ and $S_{\rm h}(\varepsilon)$ are the scattering
matrices of the normal system for  electrons and holes. 
They are related as
\begin{equation} S_{\rm h}(\varepsilon)=\mathcal{T}S_{\rm
    e}(-\varepsilon)\mathcal{T}^{-1},\label{eq:Sehrel}\end{equation}where
$\mathcal{T}=i \sigma_2 K$ is the time-reversal operator for spin-$1/2$
particles. The matrix $\sigma_2$ is the second Pauli matrix acting on the spin
degree of freedom 
and $K$ is the operator of complex conjugation. 
Relation \eqref{eq:Sehrel} reflects the fact that in the normal part the dynamics of
the holes is governed by the Hamiltonian\cite{DeGennes}
\begin{equation}H_{\rm h}=-\mathcal{T}H_{\rm e}\mathcal{T}^{-1},\end{equation}
the negative of the time reversed  electron Hamiltonian $H_{\rm e}$. 

We consider the case when the normal part is time-reversal
invariant, which imposes the self duality condition $S=\sigma_2 S^T \sigma_2$
on the scattering matrix. (The superscript $T$ refers to matrix transposition.)
The elements of $S_{\rm e}(\varepsilon)$ change significantly if $\varepsilon$
is changed on the scale of $\hbar/\tau_{\rm dw}$, therefore to leading
order in $\Delta\tau_{\rm dw}/\hbar$ one can neglect the energy dependence of $S_{\rm
  e}(\varepsilon)$, and take it at the Fermi energy, $S_{\rm
  e}(\varepsilon)\approx S_{\rm e}(0)$. Making use of the
self-duality of the scattering matrix, and introducing the usual block
structure
\begin{equation}S=\left(\begin{array}{cc}
r&t'\\t&r'
\end{array}\right),
\label{eq:Sblkstr}
\end{equation}
 the secular equation \eqref{eq:det4spect} can be simplified to\cite{Bee91}
\begin{eqnarray}
{\rm Det}\left[ \left(1-\frac{\varepsilon^{2}}{\Delta^{2}}\right)
-t^{\dagger}t
\sin^{2}\left(\frac{\phi}{2}\right)\right] =0.
\label{discrete2}
\end{eqnarray}
From this  equation follows the relation \eqref{eq:TnEnrel} between the energies and the transmission eigenvalues.

The correction of order $\Delta^2\tau_{\rm dw}/\hbar$ comes from considering the energy dependence
 of the scattering matrix to first order, \mbox{$S(\varepsilon)\approx S(0)+(dS/d\varepsilon)
 \varepsilon$.} For simplicity, we restrict ourselves here to the case 
 of two single-channel point contacts. (The extension to multichannel point contacts is given in
 App.~\ref{sec:appmultichannel}.) For single-channel point contacts the self-duality of the scattering matrix implies 
\begin{equation}r=\rho\openone_2,\ r'=\rho'\openone_2,\  t'=\sigma_2 t^T
 \sigma_2,\ t=\sqrt{T}U,\end{equation}
where $\rho,\  \rho'$ are complex numbers, $\openone_2$ is the $2\times 2$
unit matrix,  $1\ge T \ge 0$ and $U$ is a $2\times 2$ unitary matrix.
Writing the energy as $\varepsilon_0+\delta\varepsilon$ with 
\begin{equation}\varepsilon_0=\Delta \sqrt{1-T  \sin^2(\phi/2)},\end{equation}
 and keeping  terms up to linear order in the small quantities
$\delta \varepsilon={\cal O}(\Delta^2\tau_{\rm dw}/\hbar)$ and $\Delta\tau_{\rm dw}/\hbar$,  one finds the eigenvalue
 equation
\begin{multline}
{\rm Det}\left[\frac{\Delta^2}{4}\left(\sigma_2 Q^T_{11}\sigma_2-Q_{11}\right)
  \sin(\phi)\right.\\ \left. \mbox{}-\frac{\Delta^2}{4}({\rm
  Tr}\ Q) \frac{\varepsilon_0}{\Delta}\sqrt{1-\frac{\varepsilon_0^2}{\Delta^2}}\  \openone_2-\delta \varepsilon \right] =0
\label{eq:corrwQ}
\end{multline}
for the energy correction $\delta \varepsilon$.
The matrix $Q$ has the block structure
\begin{equation}
Q=\begin{pmatrix}
Q_{11}&Q_{12}\\
Q_{21}&Q_{22}
\end{pmatrix},
\end{equation}
inherited from the transmission-reflection block structure
  \eqref{eq:Sblkstr} of the scattering matrix.

 The second term in the
  determinant \eqref{eq:corrwQ} shifts both eigenvalues by the same amount
  $\delta\varepsilon_{\rm shift}$,  while the first, manifestly traceless
  term is responsible for the splitting $\pm\delta\varepsilon_{\rm split}$ of the
  doublet. We see that
  the splitting is determined by the effective Hamiltonian
\begin{equation}H_{\rm eff}=\Delta\frac{\tau_{\rm dw}\Delta}{\hbar}\ \Sigma\ 
  \sin(\phi),\label{eq:QHeff}\end{equation} 
with $\Sigma$ a traceless Hermitian $2 \times 2$ matrix having matrix elements
of order unity. This is the result of Chtchelkatchev
and Nazarov\cite{Cht03}. Our analysis gives an explicit relation\cite{Q11Q22} between the
matrix $\Sigma$ and the time delay matrix $Q$, 
\begin{equation}
\Sigma=\frac{\hbar}{4 \tau_{\rm dw}}\left(\sigma_2 Q^{T}_{11} \sigma_2-Q_{11}\right).
\label{eq:sigmaQ}\end{equation}
This is the key relation that will allow us, in the next section, to calculate
the level splitting distribution from the known properties of the time delay
matrix in a chaotic system.

\begin{figure}[tb]
  \begin{center}
    \includegraphics[width=0.95\columnwidth]{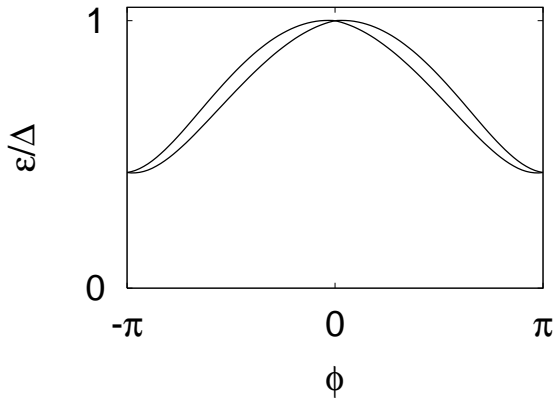} 
  \end{center}
  \caption{A schematic illustration of the splitting of the Andreev doublet as a function of the
  phase difference $\phi$ for a single-channel
  Josephson junction with spin-orbit coupling. The energies are the sum of a
  degenerate part $\varepsilon_0+ \delta\varepsilon_{\rm shift}$ that is even in
  $\phi$ and a splitting $\pm\delta\varepsilon_{\rm split} $ that is odd in
  $\phi$, as explained in the text.
  The maximal splitting is   reached at $\phi=\pi/2$.}
  \label{fig:levels}
\end{figure}

We conclude this section with a symmetry consideration. The shift
$\delta\varepsilon_{\rm shift}$ is even in $\phi$,
just like the zeroth order term $\varepsilon_{0}$. 
In contrast, the splitting $\delta\varepsilon_{\rm split}$ is odd in $\phi$.
This is in accord
with the  symmetry of the Hamiltonian $H$ that gives the full excitation
spectrum of the Josephson junction. Under time reversal, in our case of a time-reversal invariant
normal part, it 
 transforms as $\mathcal{T}H(\phi)
\mathcal{T}^{-1}=H(-\phi)$, therefore, for an eigenstate $\Psi$ one has
\begin{equation}\label{eq:Kramersevol}
\begin{array}{lll}

H(\phi)\Psi(\phi) &=& \varepsilon(\phi)\Psi(\phi), \\

H(\phi)\mathcal{T}\Psi(-\phi)&=&\varepsilon(-\phi)\mathcal{T}\Psi(-\phi).

\end{array}
\end{equation}
An Andreev doublet is therefore of the form
$\{\varepsilon(\phi),\varepsilon(-\phi)\}$. The decomposition of $\varepsilon(\phi)$ into
even and odd parts in $\phi$ amounts to a decomposition of the doublet into a degenerate
even part and an odd splitting part.  
The resulting $\phi$ dependence of the doublet is shown schematically in Fig.~\ref{fig:levels}.

\section{Splitting distribution in chaotic Josephson junctions}
\label{sec:chQdot}
As an application of our general result \eqref{eq:sigmaQ} we  calculate how the
level splittings are distributed for an ensemble of Josephson junctions where the normal
part is  a chaotic quantum dot. 
We assume that the spin-orbit coupling inside the dot is strong enough that the
dot Hamiltonian can be modeled as a member of the symplectic ensemble of RMT,
i.e. that the spin-orbit time $\tau_{\rm so}$ is
 much shorter than $\tau_{\rm
dw}$.

The splitting distribution can be obtained from the known
 distribution of the scattering matrix\cite{Bee97}, and of the
 dimensionless  symmetrized Wigner-Smith matrix\cite{Bro97a}, 
\begin{equation}Q_E = -i
\ \frac{\hbar}{\tau_{\rm dw}}\, S^{-1/2} (dS/d\varepsilon) S^{-1/2}.\end{equation}
The distributions of $S$ and $Q_E$ are
 independent\cite{Bro97a}, which makes it advantageous to express $Q$ in terms
 of  $S$ and $Q_E$:
\begin{equation}Q=\frac{\tau_{\rm dw}}{\hbar} S^{-1/2} Q_E S^{1/2}.\end{equation}
In the single-channel case one has
\begin{equation}
\begin{array}{c}
Q_E=M_1 \left(\begin{array}{cc}
1/\gamma_1\  \openone_2&0\\0&1/\gamma_2\  \openone_2
\end{array}\right) M_1^\dagger,\\
\quad \\
 S=M_2 \left(\begin{array}{cc}
e^{i\varphi_1} \openone_2&0\\0&e^{i\varphi_2} \openone_2
\end{array}\right) M_2^\dagger.
\end{array}
\label{eq:QESspecdecomp}\end{equation}
The rates $\gamma_n$  are distributed
according to\cite{Bro97a}  
\begin{equation}
  P(\gamma_1,\gamma_{2}) \propto
  |\gamma_{1} - \gamma_{2}|^{4}\
  \gamma_{1}^{4}\gamma_2^4\
  \exp[-{4(\gamma_1+\gamma_2)}]. \label{eq:Lag}
\end{equation}
 The distribution of the phases $\phi_n$
is\cite{Bee97}
\begin{equation}
  P(\phi_1,\phi_2) \propto |e^{i \phi_1} - e^{i
\phi_2}|^{4}. \label{eq:circ}
\end{equation}

The  matrices of eigenvectors $M_1$ and $M_2$ are members of the group Sp(2) of
$4\times4$ unitary symplectic matrices, and are uniformly distributed
with respect to the  Haar measure of the group\cite{Bee97,Bro97a}. The Haar measure 
is given as 
\begin{equation}d\mu \propto \sqrt{|{\rm Det} g|}\Pi_j dx_j,\end{equation}
in terms of the metric tensor $g$, defined by
\begin{equation}{\rm Tr}\left(dM dM^\dagger \right)=\sum_{ij}g_{ij}dx_i dx_j.\end{equation}
Here $\{ x_i \}$ is a set of independent variables parameterizing the Sp(2) matrix $M$.  

A convenient choice to  parameterize  Sp(2) is the decomposition
\begin{equation}M=\left(\begin{array}{cc}
\cos(\theta) &\sin(\theta)\ W\\-\sin(\theta)\ W&\cos(\theta)
\end{array}\right)\ 
\left(\begin{array}{cc}
U&0\\0&V
\end{array}\right), \end{equation}
where $W$, $U$ and $V$ are SU(2) matrices, and $\theta \in [0,\pi/2]$.
It is seen that the SU(2)$\otimes$SU(2) factor corresponding to the
block-diagonal matrix with $U$ and $V$ cancels from the spectral decomposition
\eqref{eq:QESspecdecomp} of
$Q_E$ and $S$. Using the Euler angle parameterization for SU(2),
\begin{equation}\begin{array}{c}
U\!\!=\!\!\!\!
\ 
\left(\begin{array}{cc}
e^{-i(\phi_U+\psi_U)/2}\cos(\theta_U/2) &-e^{i(\psi_U-\phi_U)/2}\sin(\theta_U/2)\\
e^{i(\phi_U-\psi_U)/2}\sin(\theta_U/2)&e^{i(\phi_U+\psi_U)/2}\cos(\theta_U/2)
\end{array}\right)\!\!,\\
\quad \\
\phi_U \in [0,2\pi],\ \psi_U \in [0,4 \pi],\ \theta_U \in [0,\pi],
\end{array}\end{equation}
and similarly for the matrices $V$, $W$, one finds that the Haar measure on Sp(2) corresponding to the chosen
parameterization is
\begin{equation}d\mu(M)\propto\sin^3(\theta)\cos^3(\theta) d\theta\!  \prod_{j=U,V,W}\!
\sin(\theta_j)d\phi_j d\theta_j d\psi_j.\end{equation}

\begin{figure}
\epsfig{file=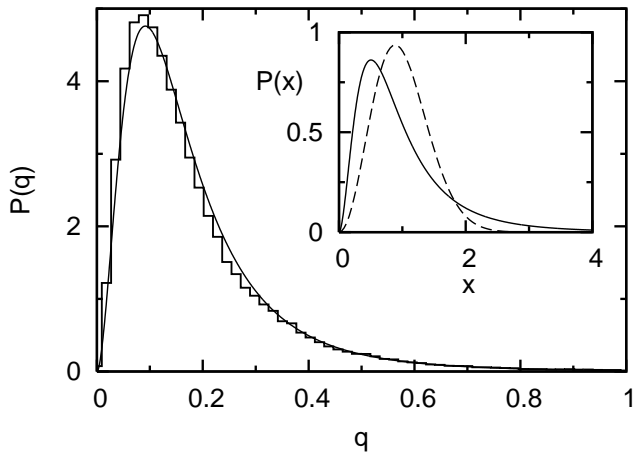,width=6.3cm,angle=270}
\caption{Main plot: Distribution of the maximal splitting of the Andreev levels
  (reached at $\phi=\pi/2$) in units of $\Delta^2 \tau_{\rm dw}/\hbar$. The smooth
  curve is the prediction of Random Matrix Theory calculated from Eq.~\eqref{eq:Pqint}, the histogram is the
  result of a numerical simulation using the spin kicked rotator. Inset:
  comparison of the Andreev doublet splitting distribution (solid line) and the Wigner
  surmise (dashed line). For this comparison, the energies are rescaled such that the mean of the
  distributions is unity. }
\label{fig:dist}
\end{figure}

We define the maximal dimensionless splitting $q$ of the Andreev levels
(reached at $\phi=\pi/2$) by the formula
\begin{equation}\delta \varepsilon_{\rm
  split}= q \Delta\frac{\Delta \tau_{\rm dw}}{\hbar}\sin(\phi).\end{equation}
The distribution of $q$ is given by
\begin{equation}
\begin{array}{c}
P(q)=\int d\mu(S) d\mu(Q_E) \delta(q-\sqrt{-\rm{Det}(\Sigma)}),\\
\quad \\
d\mu(Q_E)=d\mu(M_1) d\gamma_1 d\gamma_2 P(\gamma_1,\gamma_2),\\
\quad \\
d\mu(S)=d\mu(M_2) d\varphi_1 d\varphi_2 P(\varphi_1,\varphi_2).\\
\end{array}
\label{eq:Pqint}\end{equation}
Eq.~\eqref{eq:Pqint} can be evaluated numerically. The resulting distribution is shown in Fig.~\ref{fig:dist}.
The first two moments of $q$ are 
\begin{equation}\langle q \rangle = 0.181,\quad \sqrt{\langle q^2 \rangle-\langle q
  \rangle^2} = 0.152\ \ .\end{equation}
The splitting distribution near zero behaves as 
\begin{equation}P(q)\sim q^2\quad (q \rightarrow 0).\end{equation}
For large splittings we find
\begin{equation}P(q)\sim q^{-6}\quad (q \rightarrow \infty).\label{eq:largeasy}\end{equation}

In order to check our prediction~\eqref{eq:Pqint} for the level splitting
distribution, we have numerically simulated the chaotic quantum dot Josephson
junction  of Fig.~\ref{fig:setup}
using the  spin kicked rotator\cite{Sch89, Bar05}. The spin kicked rotator is
a dynamical model, from which one can extract scattering matrices
characteristic of chaotic cavities. These scattering matrices are given by
\begin{equation}
  \label{eq:Smatrix}
S(\varepsilon)={\cal P}[e^{-i\varepsilon}-{\mathcal F}(1 - {\cal P}^T{\cal P})]^{-1}{\mathcal F}{\cal P}^T,
\end{equation}
where $\mathcal F$ is a $2M\times 2M$ matrix giving the stroboscopic time
evolution of the model and ${\cal P}$ is a $4\times 2M$ projection matrix projecting
onto  the two single-channel point contacts (the factors of $2$ in the
dimensions are because of the spin). The
quasienergy $\varepsilon$  plays the role of the energy variable, measured in
units of $\hbar/t_0$ with $t_0$ the stroboscopic time.  For a more 
detailed description of this numerical model we refer the reader to Ref.~\onlinecite{Bar05}.

Scattering matrices generated through Eq.~\eqref{eq:Smatrix} are inserted into
the secular  Eq.~\eqref{eq:det4spect}, and the  roots are found by varying the
quasienergy. 
The dwell time in this model is $\tau_{\rm dw}=M/2$ (again in units of
$t_0$). We take $M=100$ and $\Delta=2\cdot 10^{-4}$ (in units of $\hbar/t_0$),
so that $\Delta\tau_{\rm dw}/\hbar=10^{-2}\ll 1$. By sampling about $10^5$ different ${\cal F}$, $P$, and $\phi$  we numerically obtain the distribution $P(q)$ shown in  Fig.~\ref{fig:dist} together with  the analytical
 result~\eqref{eq:Pqint}.  The agreement is very good.

\section{Discussion}
\label{sec:conclude}
\subsection{Summary}
We have investigated the effect of spin-orbit coupling on the subgap
spectrum of single-channel Josephson junctions. Using the scattering matrix
approach and considering the energy dependence of the scattering matrix to
first order we obtained a simple
relation, Eq.~\eqref{eq:sigmaQ},  between the effective Hamiltonian governing the level splitting and
the quantum mechanical time delay matrix  $Q=-i S^\dagger dS/d\varepsilon$.
This relation allowed us to find the splitting distribution for an ensemble of
chaotic Josephson junctions using the known properties of $Q$.
We verified our result numerically by simulating the chaotic Josephson junction
using the spin kicked rotator,  and we found excellent agreement.

\subsection{Comparison of the splitting distribution with the Wigner surmise}

In the inset of Fig.~\ref{fig:dist} we compare the splitting distribution of
the Andreev doublet with the Wigner surmise of RMT\cite{mehta},
\begin{equation}
P_{\rm W}(x)=\frac{32}{\pi^2}\ x^2 \ \exp\left(-\frac{4x^2}{\pi}\right). 
\label{eq:PW}\end{equation}
(For this comparison the energy scale is set such that the average splitting
is unity.) The motivation behind this comparison is the fact that the Wigner
surmise is also a splitting distribution: as shown in App.~\ref{sec:appBsplit} it
describes the distribution of the splittings of Kramers doublets for normal
chaotic quantum dots with spin-orbit coupling in the case that the
time-reversal symmetry is broken by a magnetic field. 

At small splittings, both $P$ and $P_{\rm W}$ decay quadratically. This
quadratic decay is a generic feature of the splitting of a
Kramers degenerate level due to time-reversal symmetry breaking. It
follows from the fact that the splitting Hamiltonian is a $2\times 2$
Hermitian traceless matrix without further symmetries and from a power
counting argument\cite{Haake} similar to the one leading to the
quadratic decay of $P_{\rm W}$.

While at small splittings the two distributions decay in the same way, we find
qualitative differences in the opposite limit. 
At large splittings $P$ decays like a power law in contrast to the exponential
decay of $P_{\rm
W}$ [cf.\ Eqs.\ \eqref{eq:largeasy} and \eqref{eq:PW}]. 

We attribute the deviation of $P$ from the Wigner surmise to the
nonuniform way in which time-reversal symmetry is broken: While the
magnetic field in App.\ B acts {\em uniformly\/} throughout the normal
quantum dot, the superconducting phase difference in the Josephson
junction acts {\em nonuniformly\/} at the point contacts.

\section*{ACKNOWLEDGMENTS}
 This work was supported by the Dutch Science Foundation NWO/FOM. We
also acknowledge support by the European Community's Marie Curie 
Research Training Network under contract MRTN-CT-2003-504574, Fundamentals of Nanoelectronics.

\appendix
\section{Splitting Hamiltonian for multichannel Josephson junctions}
\label{sec:appmultichannel}

We generalize the relation \eqref{eq:sigmaQ} between the splitting Hamiltonian
and the time delay matrix to the case that each of the two point contacts
supports $N/2$ propagating modes. (The single-channel case of
Sec.~\ref{sec:SQsplit} therefore corresponds to $N=2$.)
In the multichannel case, after the steps leading to Eq.~\eqref{eq:corrwQ} one
arrives at the equation
\begin{equation}{\rm Det}\left[H_0+\frac{\Delta^2}{2}\ K
-\delta \varepsilon\right] =0,\end{equation}
where
\begin{equation}H_0=\frac{\Delta^2}{2 \varepsilon_{n}^{(0)}}\left[1-\left(\frac{\varepsilon_n^{(0)}}{\Delta}\right)^2-t^{\dagger}t\sin^{2}\left(\frac{\phi}{2}\right)\right],\end{equation}
\begin{equation}\varepsilon_{n}^{(0)}=\Delta \sqrt{1-T_n \sin^2(\phi/2)},\end{equation} 
and $K$ is a matrix with elements of order $\tau_{\rm dw}/\hbar$. 
An eigenvector of $t^\dagger t$ with eigenvalue $T_n$ is also an eigenvector
of $H_0$ with zero eigenvalue. The first
order correction to the zeroth order energy $\varepsilon_{n}^{(0)}$ is the
first order perturbative correction to this zero
eigenvalue.  

We introduce the $N \times 2$ matrices $W_n$ and $W_n'$ which contain the two orthonormal eigenvectors
of, respectively,  $t^\dagger t$ and $t'^\dagger t'$, both corresponding to the
eigenvalue $T_n$. In terms of these matrices we define the matrices $q_{1n}$
and $q_{2n}$ by 
\begin{equation}q_{1n}=W_n^\dagger Q_{11} W_n, \quad q_{2n}=W_n'^\dagger Q_{22} W_n' \ .\end{equation}
We find that the shift of the Andreev doublet at $\varepsilon_n^{(0)}$ is given by
\begin{equation}
\delta\varepsilon_n^{\rm shift}=
-\frac{\Delta^2}{4}\frac{\varepsilon_{n}^{(0)}}{\Delta}\sqrt{1-\left(\varepsilon_{n}^{(0)}/\Delta\right)^2}
\left({\rm  Tr}\ q_{1n}+{\rm  Tr}\ q_{2n}\right),\end{equation}
while the splitting $\delta\varepsilon_n^{\rm split}$ is given by the two
eigenvalues of the traceless Hermitian matrix
\begin{equation}
H_{\rm eff}^{(n)}=\frac{\Delta^2}{4}\left(\sigma_2\ q_{1n}^{T}
  \sigma_2-q_{1n}\right)\sin(\phi).
\end{equation}

\section{Splitting distribution for normal chaotic quantum dots}
\label{sec:appBsplit}

We calculate the splitting distribution of a Kramers degenerate level for
normal chaotic quantum dots with spin-orbit coupling, in the case that the
time-reversal symmetry is broken by a magnetic field. 

The Hamiltonian of the system is decomposed into two parts,
\begin{equation}H=H_0+A,\quad\quad H_0^\dagger=H_0,\ A^\dagger=A,\end{equation}
where $H_0$ and $A$ are $2M \times 2M$ matrices (the factor of two is due to
the spin). They  satisfy
\begin{equation}{\cal T}H_0{\cal T}^{-1}=H_0,\ \  {\cal T}A{\cal T}^{-1}=-A.\end{equation}
The matrix $H_0$ models the time-reversal invariant part of the Hamiltonian and $A$ is a
time-reversal symmetry breaking term.

The eigenvalues of $H_0$ are doubly degenerate (Kramers degeneracy). Considering a doublet with
energy $E_0$,  with corresponding eigenvectors $u_1$, $u_2={\cal T}u_1$,
\begin{equation}H_0 u_1=E_0 u_1, \quad H_0 u_2=E_0 u_2, \end{equation}
and treating $A$ as a perturbation,  first order degenerate perturbation theory
leads to the splitting of the Kramers doublet by an amount $\pm \delta
\varepsilon_{\rm split}$. We find 
\begin{equation}\delta \varepsilon_{\rm split}= \sqrt{\langle u_1,A u_1\rangle^2+|\langle u_1,A u_2\rangle|^2}.\end{equation} 

For  chaotic billiards, the splitting distribution 
is given by\cite{mehta}
\begin{equation}P(\lambda)=\int dU\ \rho(U)\int dA\
  P(A)\delta\bigl(\lambda-\delta \varepsilon_{\rm split}\bigr),\end{equation}
where $U$ is the matrix of eigenvectors of $H_0$, distributed according to
$\rho(U)$. (The form of $\rho(U)$ is not needed for the derivation.) The matrix $A$ has
distribution
\begin{equation}P(A)\propto \exp\left(-v^2 {\rm Tr}\ A^2
    \right),\end{equation}
where $v$ is a positive number. 
Using the fact that $P(A)dA$ is invariant under a unitary transformation with
the matrix of eigenvectors of $H_0$, one finds 
\begin{equation}P(\lambda)=\int da\ db\ dc\ P(a,b,c)\delta(\lambda-\sqrt{a^2+b^2+c^2}),\label{eq:Plabc}\end{equation}
where 
\begin{equation}P(a,b,c)\propto \exp[-2  v^2 (a^2+b^2+c^2)].\end{equation}
After changing to polar coordinates the integral \eqref{eq:Plabc} can be
evaluated straightforwardly, and after rescaling from $\lambda$ to $x$, defined by $\int dx\
P(x)x=1$, one  arrives at the Wigner surmise \eqref{eq:PW}.


\begin{thebibliography}{16}
\bibitem{Q11Q22}With a little algebra one can show that the replacement of  $Q_{11}$
  with $Q_{22}$ in the definition \eqref{eq:sigmaQ} of $\Sigma$ amounts to a unitary
transformation. It is, therefore, a matter of choice to use $Q_{11}$ or $Q_{22}$
to obtain the level splitting.
\end{thebibliography}
\end{document}